\documentclass[journal=jctcce]{achemso}

\usepackage{graphicx}
\usepackage{subscript}
\usepackage{subfig}
\usepackage{xspace}
\usepackage{enumerate}
\usepackage{xcolor}
\usepackage{todonotes}
\usepackage{xr}
\usepackage{siunitx}
\usepackage[version=3]{mhchem}

\newcommand{\lm}{\ensuremath{\lambda_\mathrm{max}}\xspace}
\newcommand{\pka}{\ensuremath{\mathrm{p}K_\mathrm{a}}\xspace}

\title{Sampling the protonation states: pH-dependent UV absorption spectrum of a polypeptide dyad}

\author{Elisa Pieri}
\author{Vincent Ledentu}
\author{Nicolas Ferr\'e}
\email{nicolas.ferre@univ-amu.fr}
\affiliation{*Aix-Marseille Univ, CNRS, Institut de Chimie Radicalaire, Marseille, France}

\begin{document}

\begin{abstract}
When a chromophore interacts with titrable molecular sites, the modeling of its photophysical properties requires to take into account all their possible protonation states. We have developed a multi-scale protocol, based on constant-pH molecular dynamics simulations coupled to QM/MM excitation energy calculations, aimed at sampling both the phase space and protonation state space of a short polypeptide featuring a tyrosine--tryptophan dyad interacting with two aspartic acid residues. We show that such a protocol is accurate enough to reproduce the tyrosine UV absorption spectrum at both acidic and basic pH. Moreover, it is confirmed that UV-induced radical tryptophan is reduced thanks to an electron transfer from tyrosine, ultimately explaining the complex pH-dependent behavior of the peptide spectrum.
\end{abstract}

\section{Introduction}

The classical atomistic modeling of a biological molecule like a polypeptide, a protein or a DNA double helix usually involves a converged sampling of its configuration space, i.e. atom positions and velocities. Molecular dynamics (MD) simulations, in which trajectories are generated by solving classical Newton equations, are clearly among the most popular available methods and take benefit of continuous improvements on both the software (eg replica-exchange, accelerated MD) and hardware sides (GPUs, Anton)\cite{Durrant_11,Grand_13,Perilla_15,Mori_16}. A typical MD simulation starts with some required input parameters: the force-field defining the atom-atom interaction energy and a set of atom coordinates and velocities used as initial conditions. The latter geometrical parameters are commonly obtained from available experimentally derived structures, often by means of X-ray diffraction or NMR spectroscopies\cite{Schneider_09}. However, no information is usually found regarding the distribution of the protonation states of titrable moieties like aspartic acid, lysine side-chains and similar in a protein. Hence the model needs to be complemented by an educated guess of these protonation states.

Most of the times, the protonation state of a titratable moiety is determined by comparing its \pka with the pH of the system. Of course, experimental \pka are macroscopic values which can barely be attributed to a given titrable site in a molecular system featuring several, and possibly interacting, sites. Hence empirical methods have been developed to give a quick and rough estimation of effective microscopic \pka values for all the titrable sites in a system. For instance, the PropKa approach\cite{Olsson_11} uses an available 3-dimensional structure to estimate amino-acidic \pka values in a protein. On the other hand, the recently developed constant-pH molecular dynamics (CpHMD) method\cite{Goh_12,Swails2012,Swails2014,Huang_16a} has been especially designed to sample the protonation states of titrable amino-acids as a function of pH. Roughly speaking, this method introduces a Metropolis-based probability eventually allowing to change protonation states during the course of a normal MD simulation. This method has been shown to efficiently sample both the phase space and the protonation state space at the same time, given that sufficiently long trajectories are produced. Ultimately, the CpHMD simulations result in accurate \pka predictions\cite{Lee_14a}.

However, instead of using this information to decide on the most probable protonation states of the titrable sites, the same CpHMD trajectories can be exploited to give access to an ensemble of structures featuring a probability distribution of the protonation states at a given pH value, in agreement with the computed \pka values. Furthermore, this ensemble can be used to calculate in a second step any molecular property whose value depends on the pH. This is precisely the target of the present study, in which the pH-dependent UV absorption spectrum of a small polypeptide is simulated for the first time. In such a case, the properties of interest (vertical excitation energies and oscillator strengths) have to be evaluated by a quantum mechanical method coupled to an approximate description of the interactions between the chromophore and its environment (QM/MM)\cite{Senn_09}. To the best of our knowledge, all the QM/MM models reported in the literature assume a single and constant (i.e. most probable) distribution of the protonation states (which will be called microstate in the following). In other words, the calculated molecular property is somehow biased towards this particular microstate.

The here-proposed CpHMD-then-QM/MM work-flow can be seen as the generalization of the routine MD-then-QM/MM approach\cite{Houriez_08,Houriez_10,Olsen_10a}, used when (classical) nuclear motion contributions to a given molecular property are needed. The successful application of such a work-flow relies on a statistically meaningful selection of snapshots along the MD trajectory. Moreover the number of such snapshots has to be large enough to ensure the convergence of the property averaged value, ie with a reasonable standard deviation. In the case of the CpHMD approach, reaching such a convergence is certainly more involved than in standard MD, since the phase space is complemented with the protonation state space\cite{Swails2014}.

After having presented the peptide, we will briefly present some details regarding the CpHMD and QM/MM models we used in the present study, and finally we will describe the procedure that we followed to obtain the spectrum.

\section{Computational details}
\paragraph{The Peptide M.} The subject of our study is a $\beta$-hairpin 18-mer, named Peptide M, designed by Barry \textit{et al.}\cite{Pagba2015} and containing two UV-absorbing chromophores: tyrosine (Y5) and tryptophane (W14). The ultraviolet absorption spectrum features a dependency upon the pH value: the trace obtained by subtracting the tryptophan spectrum (recorded in water) from that of Peptide M (i.e. the Y5 spectrum) is always dominated in the 250-to-350-nm region by the $\pi-\pi^*$ transitions of the phenol ring, but the \lm undergoes a red-shift from $\sim$ 283 nm at pH 5 to $\sim$ 292 nm at pH 11. There is also an additional red-shift of a few nm at pH=5 with respect to tyrosine in water. Quoting \cite{Pagba2015}, "\emph{the red shift of the tyrosine ultraviolet spectrum in Peptide M is attributable to the close proximity of the cross-strand Y5 and W14 to form a Y5-W14 dyad.}"
\begin{figure}[h]
	\includegraphics[width=0.75\textwidth]{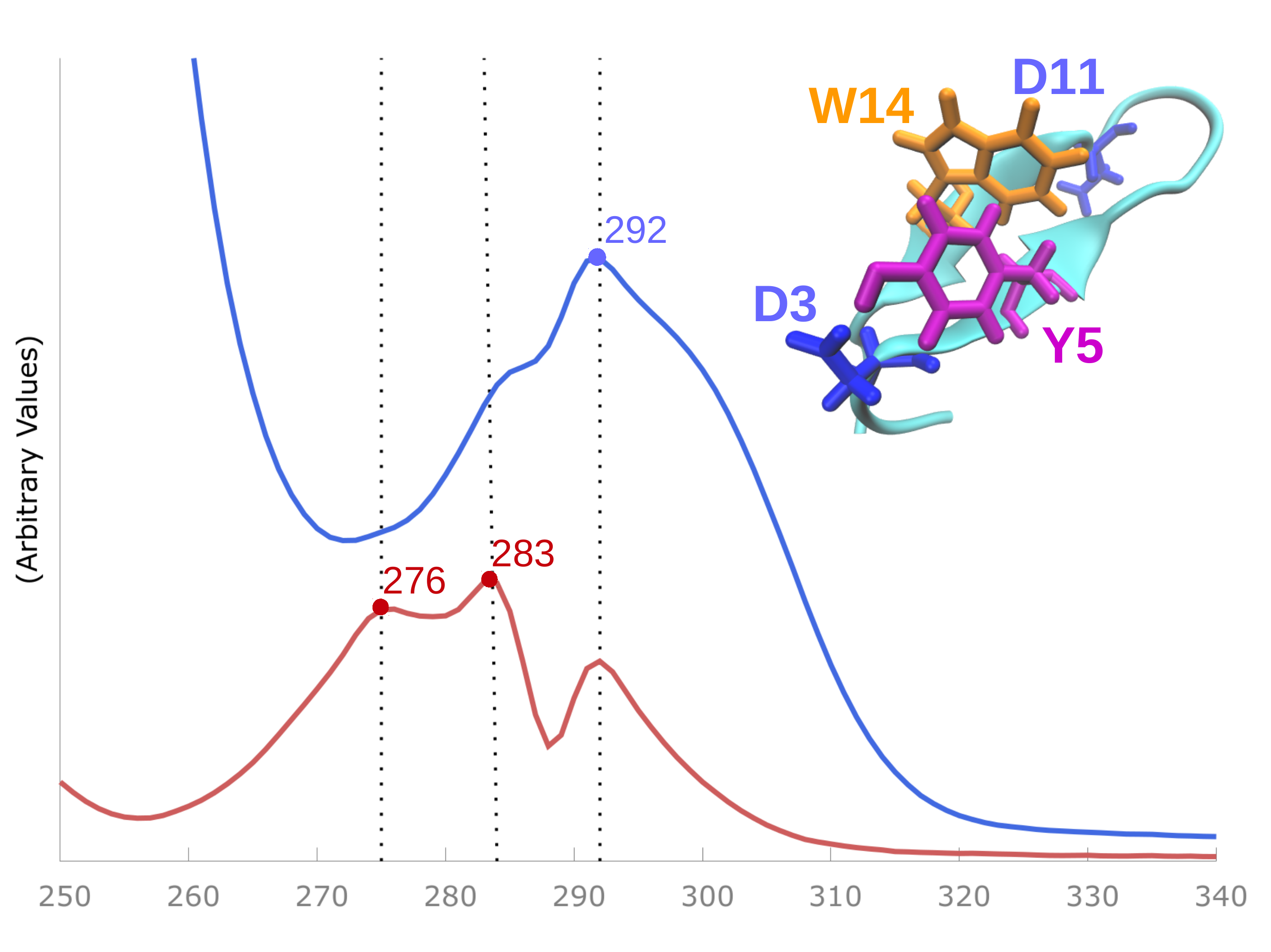}
	\caption{Peptide M structure and experimental UV absorption spectra (in nm)\cite{Pagba2015} of Y5 in Peptide M at pH=5 (red) and pH=11 (blue).}
	\label{fig:peptideM_spectrum}
\end{figure}
It is rightful, being the experimental \pka of the tyrosine side chain in water $\sim$ 10.9\cite{Cantor1980}, to attribute this behavior to the deprotonation occurring at basic pH; however, the presence of two other titratable residues, aspartic acids D3 and D11, contributes to the complexity of the protonation microstates landscape. The small size of the peptide and the limited number of titratable amino-acids make this system the ideal case study for the development and testing of our method.

\paragraph{CpHMD Method.} We carried out the CpHMD method simulations in explicit solvent using a discrete protonation state model as presented by Roitberg \textit{et al.}\cite{Swails2014} and implemented in the AMBER16 software suite\cite{Case2016}. In this method, the standard molecular dynamics steps are performed in explicit solvent, and periodically interspersed with attempts to change the protonation state in GB implicit solvent, which avoids the problem of the solvent molecule orientation; these attempts are regulated by a Metropolis Monte Carlo approach. After a successful protonation state change, which is handled by changing the charges on each atom of the residue according to the designed force field (AMBER FF10), the solvent molecules and non-structural ions are restored and relaxed, keeping the solute frozen; then, the velocities of the solute atoms are recovered, allowing the standard dynamics to continue. 
\\For this type of calculation, we made use of the replica exchange technique applied along the pH-dimension (pH-REMD), in order to enhance the sampling capabilities and get an acceptable convergence\cite{Itoh2011,Swails2012} in the given time. Our simulations were carried out using periodic boundary conditions and with a total length of 40 ns, which we considered a good compromise between accuracy in the convergence and computational time; we used 8 pH-replicas, spanning from pH 3 to pH 6 and from pH 9 to pH 12 with one pH unit as interval.
\\For the single microstates trajectories, we used the temperature replica exchange technique (T-REMD), aiming at overcoming small energy barriers and therefore exploring exhaustively the potential energy surface; we chose 6 temperature values from 260 and 360 K, with a 20 K interval.

\paragraph{QM/MM Method.} We extracted 40000 equally spaced snapshots from the trajectories at pH 5 and 11, and coupled each frame to the corresponding protonation microstate. This data allowed us to get a spatial distribution of point charges.
\\We chose the tyrosine side-chain as QM subsystem, inserting a hydrogen link-atom between $C\alpha$ and $C\beta$, and calculated the electrostatic potential acting on each QM atom using a direct sum approach over regularly spaced images of the primitive cell used in the MD simulations. $7^3$ image boxes ensure the convergence of the electrostatic potential. In the case of an electrically charged system with total charge $q_t$, we neutralized each image by placing a $-q_t$ charge at its center. In other words, the electrostatic potential experienced by the QM subsystem originates from the (charged) primitive cell and from neutralized images. Electrostatic embedding of the QM subsystem is realized thanks to the ESPF method\cite{Ferre_02}, as implemented into our local version of Gaussian09\cite{Gaussian09}.

The \lm and oscillator strengths for the first four excited states were calculated using the Gaussian09 package\cite{Gaussian09} at the TD-DFT B3LYP/6-311G* level of theory; this choice is justified by the aim of seeking qualitative and not necessarily quantitative accordance with the experimental data. Nevertheless we have assessed the quality of the TDDFT $S_0\rightarrow S_1$ vertical excitation energy with respect to state-of-the-art CASPT2 ones obtained using the Molcas package\cite{Aquilante_15} on a subset of 10 representative structures for both the protonated and deprotonated forms of Y5 side-chain. A full $\pi$ + oxygen lone pairs active space has been selected in the CASSCF calculations, using the triple-$\zeta$ ANO-L-VTZP basis set together with the resolution of identity based on the Cholesky decomposition of two-electron integrals\cite{Bostrom_10}. All details are available in Supporting Informations. Inspection of Table \ref{tab:caspt2_benchmark} clearly shows the qualitative agreement between CASPT2 and TDDFT vertical excitation energies. In particular, the ~100 nm red-shift caused to the $S_0\rightarrow S_1$ transition by tyrosine deprotonation is correctly reproduced (90 nm).
\begin{table}
	\begin{tabular}{|c|c|c|c|c|}
		\hline 
		& \multicolumn{2}{c|}{Protonated} & \multicolumn{2}{c|}{Deprotonated}  \\ 
		\hline 
		& CASPT2 & TDDFT & CASPT2 & TDDFT \\ 
		\hline 
		$S_0 \rightarrow S_1$ & 271 & 258 & 371 $\pm$ 12 & 348 \\ 
		\hline 
	\end{tabular} 
	\caption{Benchmark vertical excitation energies of tyrosine (in nm), averaged over 10 different configurations for each protonation micro-state.}
	\label{tab:caspt2_benchmark}
\end{table}

\paragraph{Spectrum Elaboration.} The absorption spectra were generated at room temperature with normalized Lorentzian functions from the excitation energies for the first four excited states and the corresponding oscillator strengths using Newton-X 2.0\cite{Barbatti2014,Barbatti2015}, which adopts the nuclear ensemble approach\cite{Barbatti2007}. Data of the experimental spectra published in \cite{Pagba2015} have been kindly provided by Prof. Barry\cite{PersonalCommunication}.

\section{Results}

\begin{figure}[h]
	\includegraphics[width=0.75\textwidth]{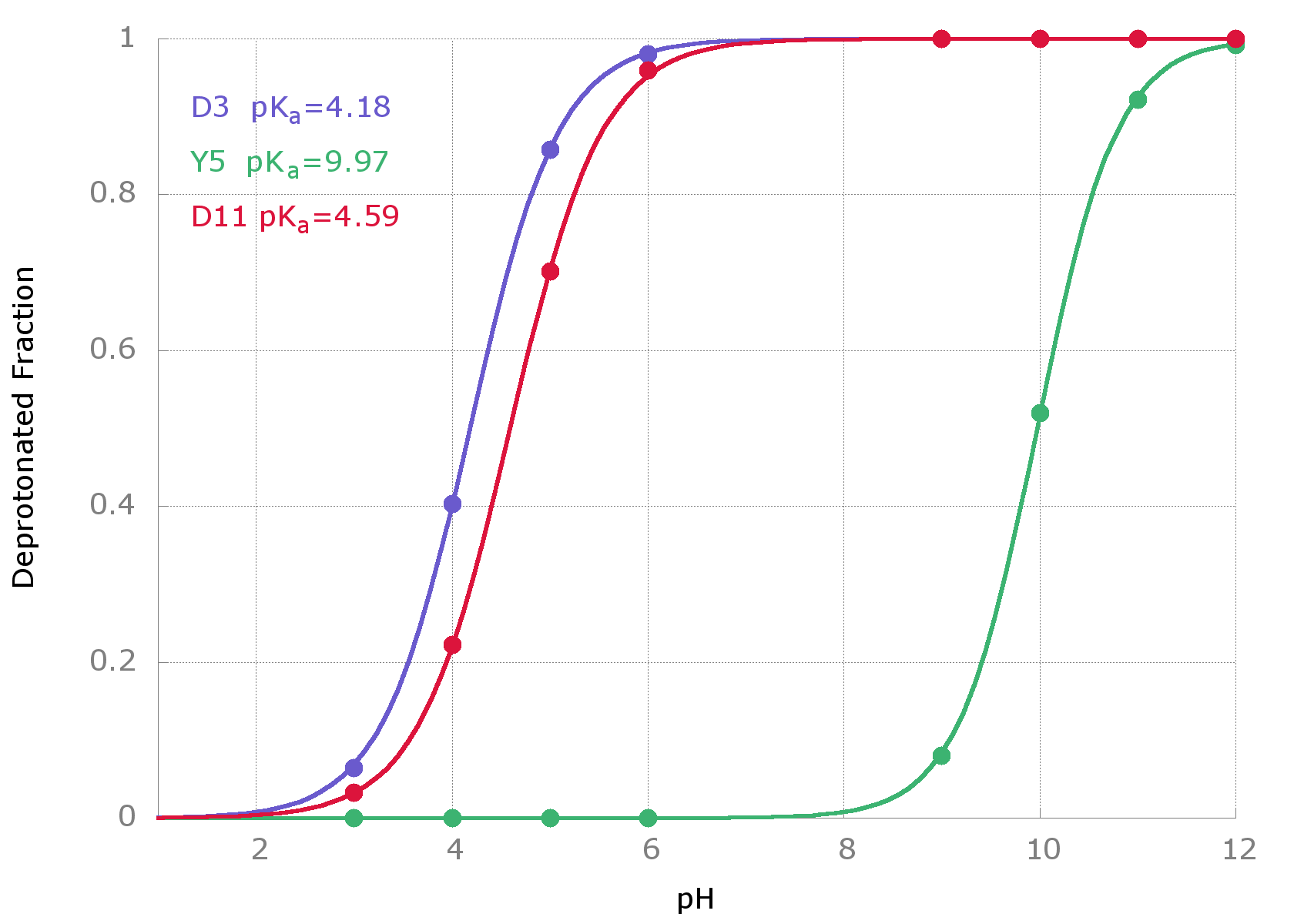}
	\caption{Titration curves (deprotonated fraction as a function of pH) from CpHMD trajectories for the three titrable residues D3, Y5 and D11 in Peptide M.}
	\label{fig:titration_curves}
\end{figure}

\paragraph{Microstate populations.}
Titration curves for all titrable residues in the system are the first useful information coming out from CpHMD simulations. Their shapes not only give qualitative informations regarding the convergence of the simulations, it also may indicate non-Henderson-Hasselbalch \cite{Onufriev_01,Po_01} behaviors possibly arising when titrable sites are strongly interacting. In the case of Peptide M (Figure \ref{fig:titration_curves}), we first produced 12 ns long trajectories for pH ranges 3--6 and 9--12. The smooth sigmoidal shape of the 3 titration curves corresponding to D3, Y5 and D11 is a good indication of a converged exploration of both the phase and protonation state spaces.
Hill fitting \cite{Onufriev_01} of the titration curves result in Hill factor $n_h$ equal to 0.96, 1.07 and 0.93 for D3, Y5 and D11 respectively. In other words, these titrable residues are interacting negligibly in the protonation state space. The subsequent analysis of the fitted curves allows to determine the microscopic \pka value of each titrable residue. As expected, the \pka value of Y5, close to 10 and not far from the reference \pka of tyrosine (in water), is higher than the aspartic acids ones (about 4). This result implies that at pH=5, the pH value at which the Peptide M absorption spectrum has been experimentally determined, the deprotonated form of both D3 and D11 dominates. Of course, at this pH value, Y5 is always protonated. On the other hand, at pH=11, corresponding to the second experimental absorption spectrum value, both D3 and D11 are always deprotonated while Y5 is predominantly in the deprotonated form.

\begin{figure}[h]
	\includegraphics[width=0.75\textwidth]{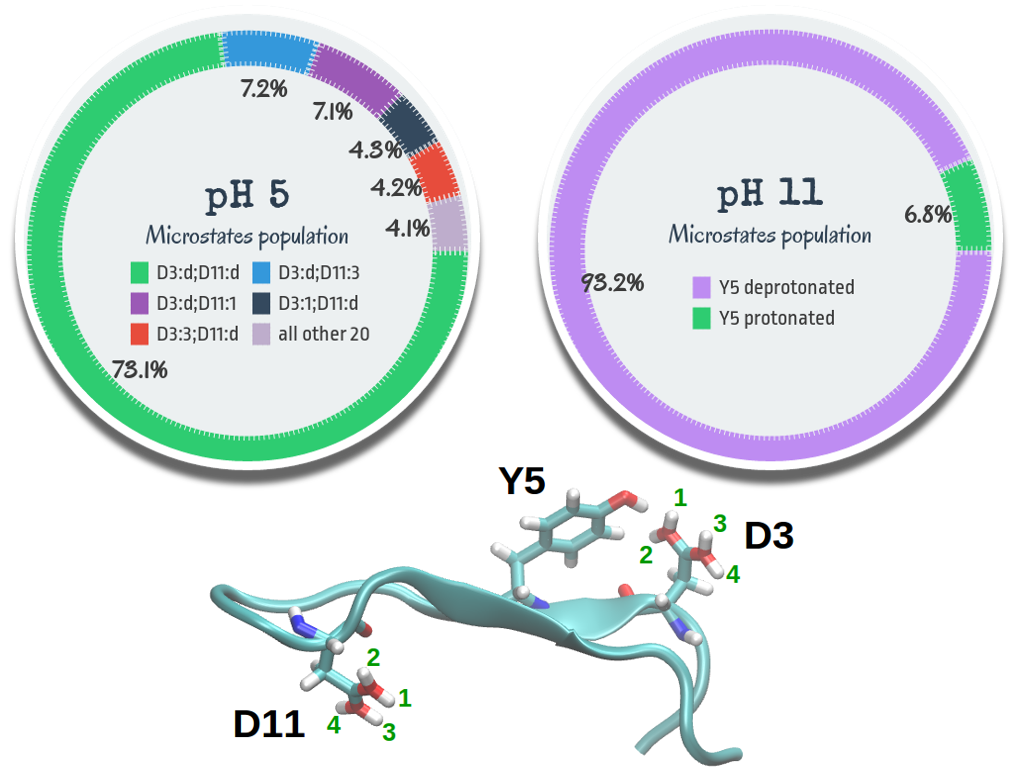}
	\caption{D3, Y5 and D11 microstate populations at pH=5 and pH=11. "d" stands for deprotonated while a number 1--4 stands for protonated in the corresponding position in Peptide M).}
	\label{fig:microstate_populations}
\end{figure}

After having established qualitatively the relative populations of the various microstates, we have expanded to 40 ns the trajectories corresponding to the same pH ranges 3--6 and 9--12. The D3 \pka value converges to 4.11, while D11 \pka is evaluated to 4.27. The detailed analysis of the microstates is reported in Figure \ref{fig:microstate_populations}. 
First, it should be noted that, in principle, the position of the acidic proton on one or the other oxygen atom of D3 (or D11) are equivalent. However, the small difference in the corresponding populations (Figure \ref{fig:microstate_populations}), about 0.1\%, indicates that the trajectories are sufficiently converged to obtain reliable population estimates. Y5 is always protonated at pH=5 while D3 and D11 are predominantly (73\%) deprotonated. On the other hand, there exists a noticeable population (14\%) in which D11 is protonated on one of the two oxygen atoms. The same occurs with D3, but to a lower extent (8\%). All other 20 possible micro-states population amounts to 4\%. 

At pH=11, both D3 and D11 are fully deprotonated, while Y5 is 93\% deprotonated. Accordingly, only 2 micro-states are populated. Compared to pH=5 populations, this situation looks easier to handle. We will see in the following that it is not necessarily true.

\begin{figure}[h]
	\includegraphics[width=0.75\textwidth]{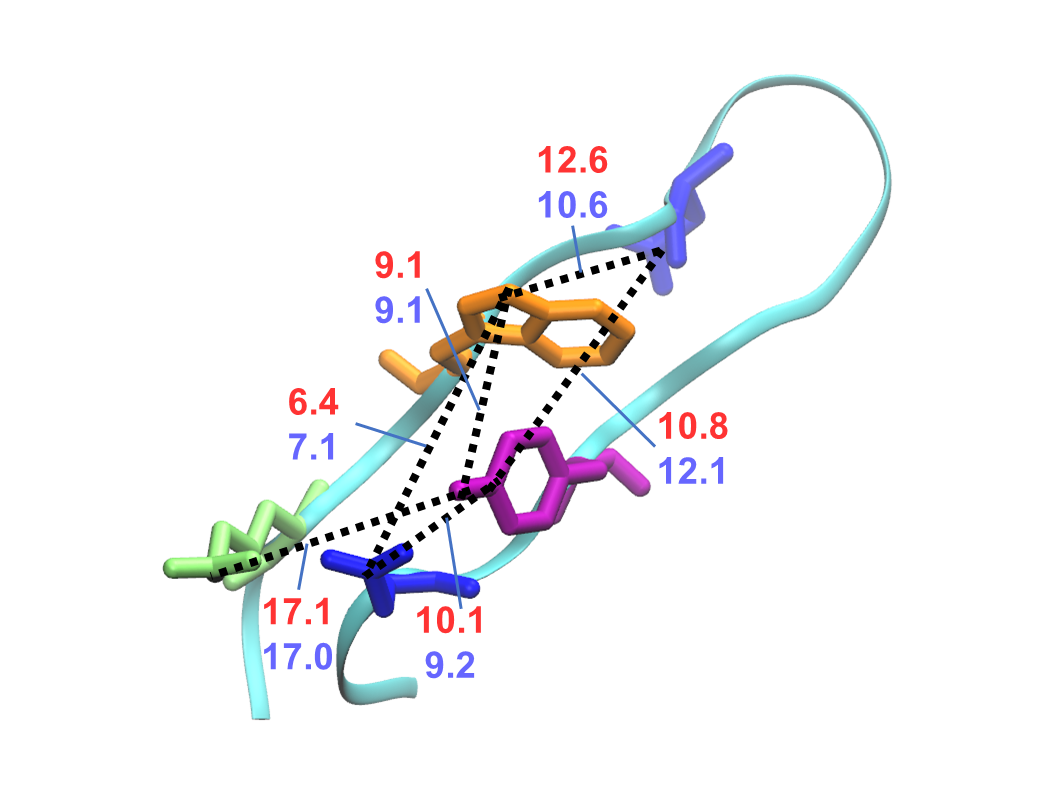}
	\caption{Selected average distances (in \AA{}) at pH=5 (in red) and pH=11 (in blue) between D3 (blue, bottom left), Y5 (purple), D11 (blue, top right), W14 (orange) and R16 (green).}
	\label{fig:distances}
\end{figure}

With respect to the \pka value of isolated aspartic acid in water (3.9), D3 closer \pka value implies a slightly larger stabilization of D11 protonated form which may be attributed to enhanced interactions between D11 and other components of Peptide M. Figure \ref{fig:distances} (complemented with Table \ref{tab:distances}) reports a selected set of average distances at pH=5 and pH=11.
\begin{table}[h]
	\caption{Selected average distances and standard deviations (in \AA{}) at pH=5 (also decomposed according to the most important microstates, see Figure \ref{fig:microstate_populations} for notation) and pH=11 between D3 (\ce{C_{\gamma}}), Y5 (\ce{O}) , D11 (\ce{C_{\gamma}}), W14 (\ce{N_{\epsilon}}) and R16 (\ce{N_{\epsilon}}).}
	\begin{footnotesize}
	\begin{tabular}{rcccccc}
	& Y5 $\cdots$ W14 & Y5 $\cdots$ D3  & Y5 $\cdots$ D11 & W14 $\cdots$ D3 & W14 $\cdots$ D11 & Y5 $\cdots$ R16 \\
	\hline
	pH=5       & $9.1 \pm 2.2$ & $10.1 \pm 2.1$ & $10.8 \pm 1.3$ & $6.4 \pm 2.9$ & $12.6 \pm 1.6$ & $17.1 \pm 2.3$ \\
	D3:d,D11:d & $8.8 \pm 0.9$ & $9.7  \pm 1.9$ & $10.7 \pm 0.8$ & $6.8 \pm 1.9$ & $10.5 \pm 0.7$ & $19.2 \pm 2.1$ \\
	D3:d,D11:3 & $11.1 \pm 1.0$ & $10.4 \pm 1.0$ & $11.5 \pm 0.9$ & $4.5 \pm 1.7$ & $13.6 \pm 1.5$ & $18.3 \pm 0.9$ \\
	D3:d,D11:1 & $8.5 \pm 0.7$ & $10.3 \pm 0.8$ & $11.5 \pm 0.8$ & $5.9 \pm 1.2$ & $11.1 \pm 0.8$ & $15.0 \pm 1.5$ \\
	D3:1,D11:d & $8.5 \pm 0.5$ & $10.7 \pm 0.7$ & $10.7 \pm 0.7$ & $6.6 \pm 1.3$ & $10.6 \pm 0.6$ & $20.4 \pm 1.6$ \\
	D3:3,D11:d & $8.0 \pm 0.7$ & $10.3 \pm 1.4$ & $11.7 \pm 0.8$ & $5.5 \pm 1.4$ & $10.4 \pm 0.7$ & $17.5 \pm 1.9$ \\
	\hline
	pH=11      & $9.1 \pm 1.6$ & $ 9.2 \pm 1.7$ & $12.1 \pm 1.0$ & $7.1 \pm 3.1$ & $10.6 \pm 1.4$ & $17.0 \pm 2.5$ \\
	\hline
	\end{tabular}
	\end{footnotesize}
	\label{tab:distances}
\end{table}
First, it should be noted that the pH does not seem to modify the average distance (9.1 \AA) between Y5 and W14. However, the corresponding fluctuations are larger at acidic pH than at basic one.
Regarding the distances between the two aspartic acids (D3 and D11) and the members of the dyad (Y5 and W14), they show different behaviors with respect to the pH. While Y5 is always closer to D3 than D11, the distance between Y5 and D3 decreases with increasing pH while the distance between Y5 and D11 increases at the same time. On the other hand, W14 is always much closer to D3 than D11. When going to acidic to basic pH, the distance between W14 and D3 slightly increases while the the distance between W14 and D11 decreases by 2 \AA.
Finally, at variance with results indicated by Pagba et al \cite{Pagba2015}, our simulations do not show evidence of strong (hydrogen-bond) interactions between Y5 and R16, the corresponding distance being always larger than 17 \AA.

\paragraph{Convergence and correlations.}
The accuracy of the proposed simulation protocol ultimately depends on the quality of the underlying statistics, i.e. the production of a sufficiently large number of uncorrelated snapshots extracted from the CpHMD trajectories. We first investigated the dependence of the \pka predicted values with the length of the trajectories, using four 10 ns and two 20 ns windows extracted from the available 40 ns trajectories at each pH value and compared them with the \pka values obtained from the 40 ns trajectories. As reported in Supplementary Informations (Figure \ref{fig:D3pKaEVO}), the D3 \pka value does not change much, converging to 4.11. However, D11 \pka value is less stable, ranging from 4.66 to 3.80 if only 10 ns of trajectory are used (Supplementary Informations, Figure \ref{fig:D11pKaEVO}). Nevertheless, the \pka value obtained from the 40 ns trajectories is 4.27.

We then determined the minimum time step between two consecutive uncorrelated snapshots extracted from the CpHMD trajectories. This was achieved by analyzing the autocorrelation function of the QM/MM vertical excitation energies (ground to the first 4 excited states) computed from 10000 snapshots separated by 1 fs. As apparent in Supplementary Informations (Figure \ref{fig:ACF}), two consecutive snapshots are uncorrelated if they are separated by about 700 fs. Because the QM/MM calculations are somehow expensive, we decided to sample the CpHMD trajectories each ps in the following.

\begin{figure}[h]
	\includegraphics[width=1.0\textwidth]{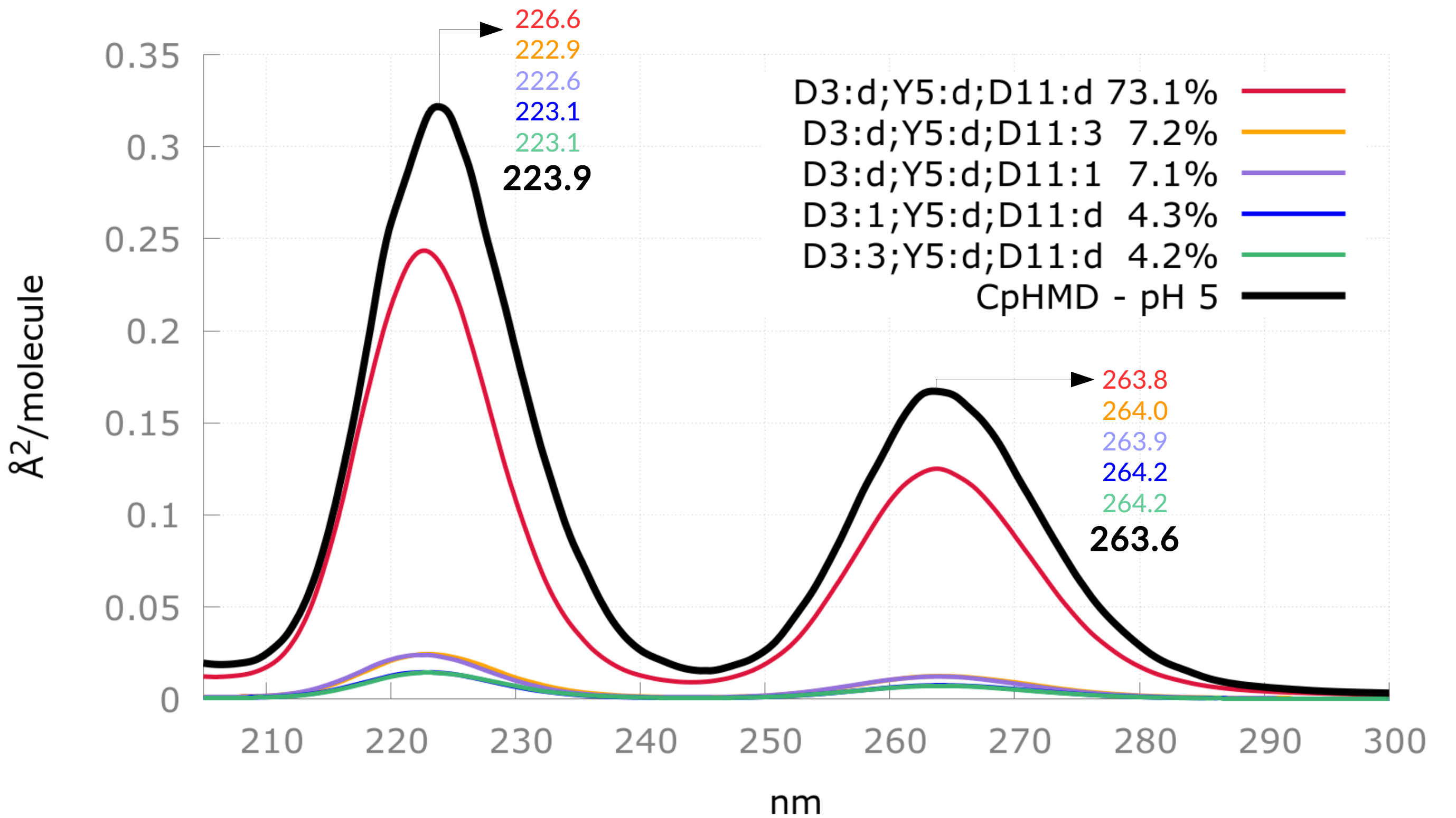}
	\caption{Computed UV absorption spectrum for Y5 in Peptide M and contributions from each important microstate at pH=5 weighted by their respective population.}
	\label{fig:acid_spectra}
\end{figure}

\paragraph{Analyzing Y5 spectrum at pH=5.}
The computed UV absorption spectrum of Y5 in Peptide M at pH=5 is reported in Figure \ref{fig:acid_spectra}. 
It includes 2 bands between 200 and 300~nm, with \lm values equal to 263.6 and 223.9~nm. Given the TDDFT 20 nm blue-shift already documented (Table \ref{tab:caspt2_benchmark}), the first \lm value may reproduce the 276 or 283~nm experimental value. The second maximum is even in better agreement with the experimental \lm at 227~nm.
\\In order to disentangle the contribution of each microstate, we have performed single microstates MD simulations with T-REMD, i.e. with a fixed distribution of protonation states, and calculated the corresponding UV spectra (Figure \ref{fig:acid_spectra}). Obviously, the main contributions originate from the most abundant microstate in which D3 and D11 are deprotonated, while Y5 is protonated. 
Accordingly, the pH=5 spectrum of Y5 could be satisfactorily modeled using this single microstate. However, it is interesting to have a look to the other contributions, i.e. the microstates in which either D3 or D11 are protonated. When considering the first absorption band, D3 protonation is responsible for 0.4~nm red-shift. On the other hand, D11 protonation comes together with a more limited 0.1 or 0.2~nm red-shift, depending on which D11 oxygen atom the proton is located on. Again, for the second absorption band, D3 causes a larger \lm red-shift (0.5~nm) than D11 (0.0 to 0.3~nm).

Such a D3 vs D11 difference, while remaining small, demonstrates that the present CpHMD-then-QM/MM protocol is able to capture subtle \lm changes caused by modifications of the protonation states. As already seen from the average distances (Figure \ref{fig:distances}), the origin of \lm slight changes cannot be simply related to the distances between Y5 and each aspartic acid. Indeed, the distance between Y5 and D3 is reduced when D3 is protonated, while it is strongly enlarged (by more than 2\AA{}) when D11 is protonated in position 3 (see Table \ref{tab:distances}).

However, we expected the final CpHMD spectrum at pH 5 (Figure \ref{fig:acid_spectra} in black) to feature \lm values very close to the ones of the most abundant microstate (Figure \ref{fig:acid_spectra} in red), while our results show a small difference (0.2~nm blue-shift) in the first absorption band and a large one (1.3~nm red-shift) in the second band. Actually, these differences cannot be justified only by the contributions of the other microstates. Indeed, their corresponding \lm are always red-shifted with respect to the main microstate \lm. While these shifts are going in the right direction in the case of the second band (222.6~nm $\rightarrow$ 223.9~nm), this is no longer true in the first band (263.8~nm $\rightarrow$ 263.6~nm). Since the distances analysis of the single microstates T-REMD simulations (Table \ref{tab:distances}) looks consistent, we conclude that the phase space sampled with the pH-REMD algorithm could possibly be less converged than the T-REMD one for this biological system. In other words, the absorption spectrum of peptide M may require more sampling than its pKa values.

\begin{figure}[h]
	\includegraphics[width=0.9\textwidth]{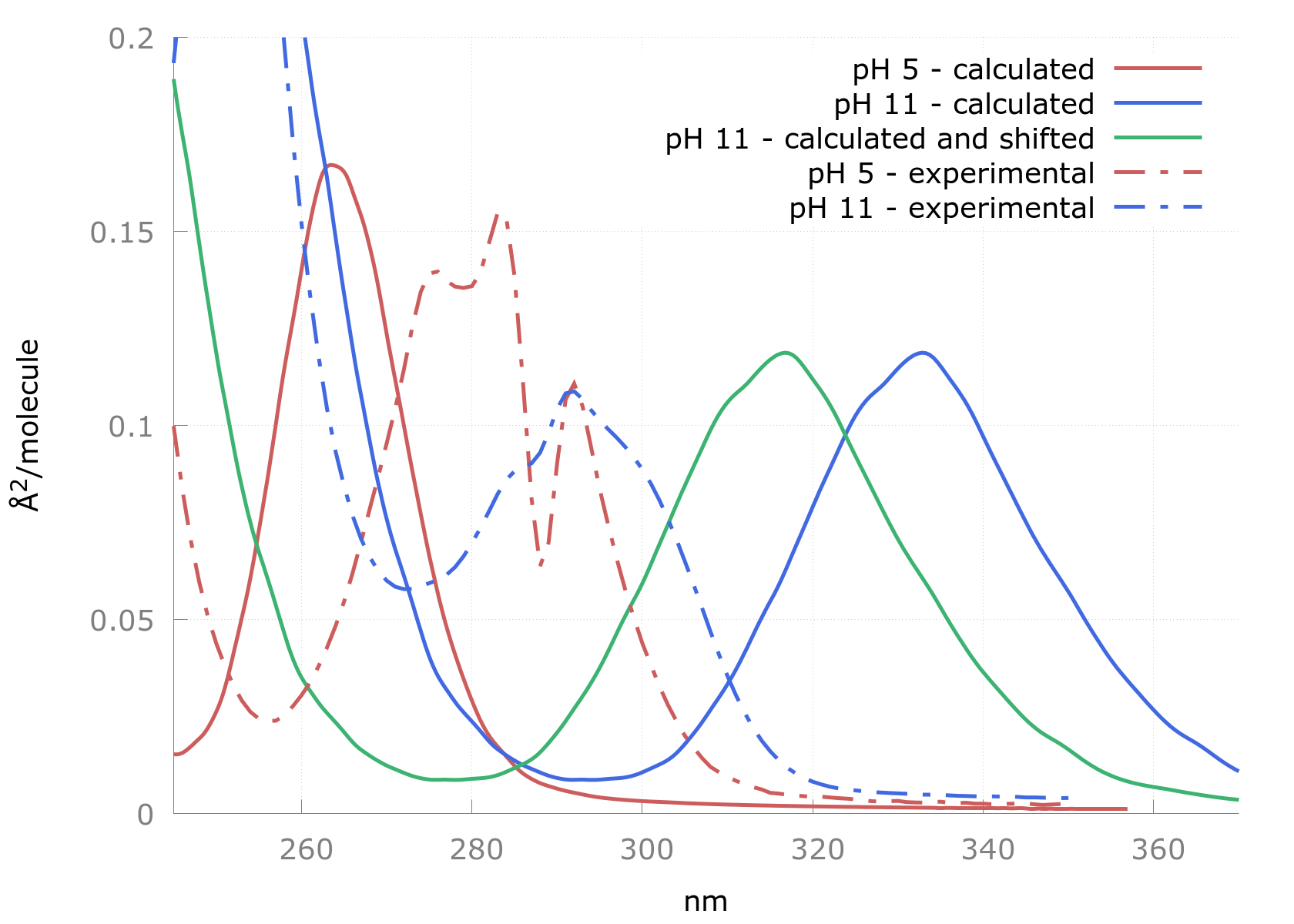}
	\caption{Y5 spectra in Peptide M at pH=5 and pH=11 (anionic Y5), both calculated and experimental\cite{Pagba2015}.}
	\label{fig:wrong_spectra}
\end{figure}

\paragraph{Comparing Y5 spectra at pH=5 and pH=11.}
The computed UV absorption spectra of Y5 in Peptide M at pH=5 and pH=11 are reported in Figure \ref{fig:wrong_spectra}, together with the experimental spectra reproduced from \cite{Pagba2015}. Going from pH=5 to pH=11, the experimentally reported red-shift is reproduced by our calculations. However, instead of a 10 nm displacement of the first absorption band, applying the CpHMD-then-QM/MM procedure results in a much larger 70 nm red-shift.
Several possible reasons may cause such a discrepancy. The use of the B3LYP/6-311G* level of theory results in the usual blue-shift of the $S_0\rightarrow S_1$ excitation energy. However, this shift does not change going from pH=5 (protonated Y5) to pH=11 (deprotonated Y5).

The quality of the QM/MM electrostatic interactions between Y5 and the surrounding water molecules may be responsible. When deprotonated, the Y5 electron density is strongly localized in the phenolate moiety, hence in close contact with the MM point charges which may in turn induce an overpolarization of Y5 electron density. As a matter of fact, we have compared QM/MM and QM-only vertical excitation energies obtained from 10 selected snapshots, treating the closest water molecules (within a 3 \AA{} distance from Y5 oxygen atom) either with point charges or quantum-mechanically. Inspection of Table \ref{tab:water_interactions} shows that i) the QM/MM approach works well when Y5 is protonated, ii) an average 16 nm blue-shift has to be applied when Y5 is deprotonated. 
\begin{table}[h]
	\caption{Influence of the level of theory (fully QM or QM polarized by water point charges (QM+q)) describing the Y5 -- water interactions on the first vertical transition (averaged over 10 snapshots, in nm).}
	\begin{tabular}{cccc}
		                &   QM  &  QM+q & $\Delta$ \\
		Protonated Y5   &  254  &  259  &    5    \\ 
		Deprotonated Y5 &  300  &  316  &   16    \\
	\end{tabular}
	\label{tab:water_interactions}
\end{table}
The corresponding shifted spectrum has been added to Figure \ref{fig:wrong_spectra}, in green. Still, a too large pH=5 to pH=11 spectral shift (about 50 nm, instead of 10 nm) is obtained.

\begin{figure}[h]
	\includegraphics[width=0.9\textwidth]{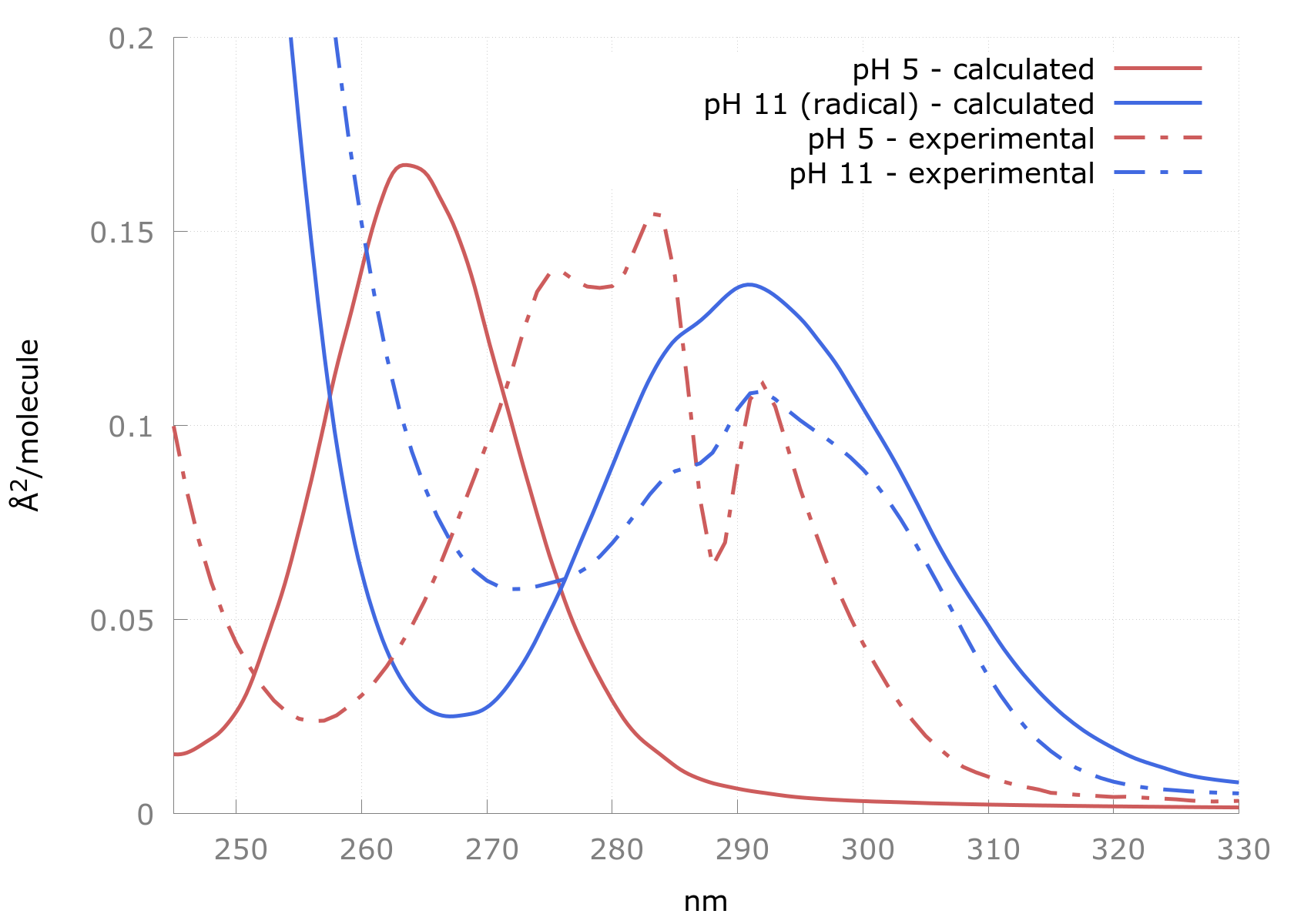}
	\caption{Y5 spectra in Peptide M at pH=5 and pH=11 (radical Y5), both calculated and experimental\cite{Pagba2015}.}
	\label{fig:correct_spectra}
\end{figure}

In their report, Barry et al \cite{Pagba2015} indicate that Y5 and W14 are actually strongly interacting: "\emph{the red shift of the tyrosine ultraviolet spectrum in Peptide M is attributable to the close proximity of the cross-strand Y5 and W14 to form a Y5-W14 dyad.}" As a matter of fact, the UV absorption spectrum of Y5 is perturbed at pH=5, but not at pH=11, with respect to the reference spectrum in water. Conversely, the UV absorption of W14 is perturbed at pH=11, but not at pH=5. Together with other spectroscopic arguments, these perturbations are interpreted as the signature of the formation of a dyad. Actually, their close proximity may promote a photoinduced electron transfer between tyrosine and transient radical tryptophan \cite{Morozova_03,Reece_05}. Assuming that W14 can be oxidized by the UV laser used in the experiment\cite{Pagba2015} to form the radical form we will denote \ce{W^{.+}} in the following, it can then react with protonated Y5 (denoted \ce{Y-OH} in chemical reaction \ref{ce1}) at pH=5 or deprotonated Y5 (denoted \ce{Y-O-} in chemical reaction \ref{ce2}) at pH=11.
\begin{align}
	\ce{W^{.+} + Y-OH &<=> W + Y-O^{.} + H+ \label{ce1} \\
	    W^{.+} + Y-O- &<=> W + Y-O^{.}} \label{ce2}
\end{align}
Whatever the pH value, these chemical reactions result in the generation of radical tyrosine which spectroscopic signature may be significantly different not only from the neutral (protonated) Y5 one, but also from the anionic (deprotonated) Y5 one. In order to test this hypothesis, we have re-analyzed the pH=11 trajectory, assuming that each deprotonated extracted snapshot (93\% of the population) actually features a radical Y5. The corresponding absorption spectrum is reported in Figure \ref{fig:correct_spectra}.
The improvement of the pH=11 spectrum is spectacular: the two experimental peaks at 283 and 292 nm are perfectly reproduced. Moreover, their respective intensities are also in agreement with experiment. From these results, we can conclude that indeed Y5 and W14 may form a dyad featuring a deprotonated radical tyrosine, possibly in equilibrium with its neutral protonated form at pH=5, even if the average distance between Y5 oxygen atom and W14 nitrogen atom in our simulations is rather large at both pH values (9.1 \AA{}). This distance could not be representative of the actual distance in the presence of the Y5-W14 dyad, when W14 is a radical species.

\section{Conclusions}

In this article, we have reported a new multi-scale protocol developed for simulating the pH-dependent photophysical properties of a peptide featuring a tyrosine-tryptophan dyad in interaction with two titrable aspartic acid residues. The modeling work-flow features two main steps: (i) the sampling of both the phase space and the protonation state space of the peptide by CpHMD simulations, (ii) the calculation of the tyrosine UV absorption spectrum by means of QM/MM calculations.

Using the replica-exchange approach, CpHMD-based \pka values of the three titrable residues are converged in tenths of ns, with uncorrelated snapshots separated by 1 ps. Using the ESPF method, the QM/MM calculations can be achieved on thousands of protonated or deprotonated tyrosine side-chains polarized electrostatically by their environment (peptide and water molecules).

At pH=5, tyrosine in Peptide M is fully protonated (neutral). However its interaction with aspartic acid or aspartate residues in various minor microstates induces small deviations from the principal microstate.

At pH=11, tyrosine in Peptide M is mostly deprotonated (ionized) while interacting with deprotonated aspartate residues. However, its experimental UV absorption spectrum cannot be explained without assuming that (i) tryptophan can be ionized by the UV-light source and (ii) radical tryptophan is reduced by an electron transferred from tyrosine which UV spectrum signature reflects its radical nature, ultimately confirming the existence of the tryptophan--tyrosine dyad.

In principle, the reported modeling protocol can be applied to the calculation of any pH-dependent molecular property, especially when it depends on a larger protonation state space, as it is the case in proteins which may feature a very large number of titrable residues.

\begin{acknowledgement}
The authors thank the French Agence Nationale de la Recherche for funding (grant ANR-14-CE35-0015-02, project FEMTO-ASR). M\'esocentre of Aix-Marseille Universit\'e and GENCI (CINES Grant 2017-A0010710063) are acknowledged for allocated HPC resources.
	
\end{acknowledgement}

\begin{suppinfo}
Convergence of the \pka values with the CpHMD simulation time. Autocorrelation function of vertical excitation energies.
\end{suppinfo}

\bibliographystyle{achemso}
\bibliography{peptideM}

\end{document}